\documentclass[aps,pre,reprint,showpacs,bibnotes,twoside,eqsecnum]{revtex4-1}

\usepackage{amsmath}
\usepackage{amssymb}
\usepackage{epsfig}
\usepackage{bm}
\usepackage{latexsym}
\usepackage{color}
\usepackage{mathtools}
\usepackage{epstopdf}
\usepackage{appendix}

\begin{document}

\title{Localization properties of harmonic chains with correlated mass and spring disorder: Analytical approach}

\author{I.~F.~Herrera-Gonz\'alez}
\affiliation{Decanato de ingenier\'ias, UPAEP University, 21 Sur 1103, Barrio Santiago, Puebla, Pue., M\'exico}
\author{J.~A.~M\'endez-Berm\'udez }
\affiliation{Instituto de F\'isica, Benem\'erita Universidad Aut\'onoma de Puebla, Apartado postal J-48, Puebla 72570, M\'exico}


\date{\today}

\begin{abstract}
We study the localization properties of normal modes in harmonic chains with  mass and spring weak disorder. Using a perturbative approach, an expression for the localization length $L_{\text{loc}}$ is obtained, which is valid for arbitrary correlations of the disorder (mass disorder correlations, spring disorder correlations, and mass-spring disorder correlations are allowed), and for practically the whole frequency band. In addition, we show how to generate effective mobility edges by the use of disorder with long range self-correlations and cross-correlations. The transport of phonons is also analyzed showing effective transparent windows that can be manipulated through the disorder correlations even for relative short chain sizes. These results are connected to the problem of heat conduction in the harmonic chain; indeed, we discuss the size scaling of the thermal conductivity from the perturbative expression of $L_{\text{loc}}$. Our results may have applications in modulating thermal transport, particularly in the design of thermal filters or in manufacturing high-thermal-conductivity materials.   
\end{abstract}


\maketitle

\section{Introduction}
Wave propagation in disordered systems has been extensively studied over the last decades since the discovery of the phenomenon of Anderson localization which inhibits wave transport in disordered media. This phenomenon was first studied in electronic systems \cite{A58}, but later it was also discovered in other kinds of media \cite{LGP87,S91,MS08}.
     
In the case of one dimensional systems with uncorrelated disorder all the electronic states become exponentially localized for an infinite system size \cite{MI70,I73}. One way of measuring the spatial extension of these states is through the quantity known as localization length. Indeed, all eigenstates are also localized in the finite system of size $N$ if $N \gg L_{\text{loc}}$ which is identified as the localized regime.        

For the  harmonic chain with no correlations of the mass (or spring) disorder, all the vibrational modes becomes exponentially localized, with the exception of the zero-frequency mode, in the thermodynamic limit. This means that for a finite system size, there is a finite fraction of vibrational modes near the zero-frequency mode which are extended regardless on the system size. This prevents the system to become a thermal insulator in contrast to what happens for electronic states.

Breaking Anderson localization is not only important in fundamental physics, but also for technological applications. One way of producing such effect in harmonic chains is by the introduction of correlated-mass or correlated-spring disorder: specific short-range correlations of the disorder can produce extended vibrational modes for some non-zero discrete frequencies when dimeric correlations are taken into account \cite{DK94,DMS93}. Whereas for some disordered models with specific long-range correlation, numerical analysis or a second order perturbative approach show that extended modes within finite frequency intervals can be produced \cite{MCRL03,HIT10,JLM15}. A similar conclusion can be obtained for electronic systems with correlated-diagonal disorder, see Ref. \cite{IKM12}, and references therein. In addition, effective mobility edges can also be produced in the Kroning-Penney model with correlated compositional and structural disorder \cite{HITT10}. Moreover, recently, it has been discovered a one dimensional tight-binding model with exact mobility edges \cite{WXZYCYZL2020}. The experimental confirmation of the existence of effective mobility edges is reported in Refs.~\cite{KIKS2001,DKHT12}. 

To understand and control the breaking of Anderson localization in harmonic chains, an analytical formula for the  localization length for the whole frequency band is required. Nevertheless, most of the expressions for $L_{\text{loc}}$ are only valid in the low frequency limit. Indeed, there exist asymptotic expressions for $L_{\text{loc}}$ for the case of purely  mass disorder with no correlations \cite{MI70} and for any kind of stationary correlations \cite{HIT15,ZLW15}, for some kinds of purely bond disorder with no correlations \cite{A83,AABOI19}, and for the case when both  mass disorder and weak bond  disorder with no correlations are present \cite{F21}. The only context in which an expression of $L_{\text{loc}}$ is obtained for practically the whole frequency band corresponds to the case of weak mass disorder with stationary correlations \cite{HIT10}.

Additionally, an expression for $L_{\text{loc}}$ allows to study analytically some aspects of the heat conduction through the disordered harmonic chain when it is embedded between two thermal baths. One of the most important quantities is the thermal conductivity defined in this context as \cite{LLP03} 
\begin{equation}
\kappa=\frac{JN}{\Delta T},
\label{Fou}
\end{equation}
with $J$ and $\Delta T$ the stationary heat flux through the harmonic chain and the temperature difference between the two reservoirs, respectively. Specifically, expressions for $L_{\text{loc}}$  have allowed to estimate the size scaling of the thermal conductivity  for uncorrelated mass disorder \cite{MI70,CL71,V79}, for mass disorder with arbitrary correlations \cite{HIT15,HM20}, and for uncorrelated spring disorder \cite{AOI18,AABOI19}. 

The importance of studying the thermal conductivity relies on the fact that it appears in the (phenomenological) Fourier's law which relates the heat current to the temperature gradient in materials that are close enough to the global equilibrium. In Fourier's law, $\kappa$  does not depend on the system size and the temperature profile is linear. Nevertheless, when one tries to explain Fourier's law from first principles in low dimensional systems one finds that, for momentum-conserving 1D systems, $\kappa$ depends on the chain size $N$  \cite{D08,L16}.  For example, in the case of harmonic chains with uncorrelated mass disorder and free boundary conditions, $\kappa$  diverges in the thermodynamic limit \cite{MI70}. Even though there exist disordered harmonic chain models where the total momentum is conserved and $\kappa$ scales normally \cite{HIT15,AOI18}, it has been proved recently that disordered harmonic chains cannot exhibit a linear temperature profile, and therefore, even if $\kappa$ becomes an intensive quantity, Fourier's law  is not satisfied \cite{HK21}. 
 
Finally, much less attention has been paid to the role that correlations between mass and spring disorder plays on the transport of phonons and heat conduction. This role is of extreme importance since in a real system there is a natural relation between the force constant and the mass, and therefore, correlations between both random variables cannot be neglected. Indeed, cooperative effects between both kinds of randomness induce the appearance of transparent states in binary alloy models \cite{ZZCRKL19} or the localization enhancement of modes when second-neighbor springs are also taking into account \cite{ASML2015}. 

In the present study we analyzed the localization and transport properties of a harmonic chain with weak mass and spring disorder. Self and cross correlations among the disordered variables are allowed and only nearest-neighbor interactions are taken into account. The paper is organized as follows. In Sec. \ref{model} the model under study is defined. In Sec. \ref{Ham} the problem is transformed into a classical two dimensional map where a uniform phase distribution arises. In Sec. \ref{locS}, the analytical solution for the localization length is obtained with a perturbative  approach. In Sec. \ref{Num} numerical simulations are performed to test our analytical results and to analyze transport properties of phonons. Here we also show how to create effective mobility edges. In Sec. \ref{conducs}, it is discussed how to obtain the size scaling dependence of $\kappa$ from the expression of $L_{\text{loc}}$, in this general context of disordered harmonic chains with correlations, and the role of the effective mobility edges on the heat conduction is also addressed. The conclusions are drawn in Sec. \ref{con}.

\section{The model}
\label{model}

We consider the infinite-size harmonic chain with weak mass and spring disorder whose dynamical equations are 
\begin{equation}
m_n\frac{d^2u_n}{dt^2}=k_n[u_n-u_{n-1}]+k_{n+1}[u_{n+1}-u_n],
\label{dynamic}
\end{equation} 
where $u_n$ is the displacement of the $n$th mass from its equilibrium position $na$, with $a$ the distance between two consecutive equilibrium positions, and $k_n$ the coupling strength between the masses $m_{n-1}$ and $m_n$. $m_n$ and $k_n$ are random variables, whose firsts moment are denoted by $M=\langle m_n\rangle$ and $K=\langle k_n\rangle$, respectively; $\langle \cdot \rangle$ represents disorder average. Whereas, their normalized variances are defined as
\begin{eqnarray}
\widetilde{\sigma}^2_m\equiv \frac{\langle m^2_n\rangle -M^2}{M^2} \ \ \text{and} \ \ \widetilde{\sigma}^2_k\equiv \frac{\langle k^2_n\rangle -K^2}{K^2}
\end{eqnarray}
with the following characteristics:
\begin{eqnarray}
4\sin^2\left( \frac{\mu a}{2}\right)\widetilde{\sigma}_m \ll 1 \ \ \text{and} \ \ \widetilde{\sigma}_k \ll 1.
\label{weak}
\end{eqnarray}
The first expression represents an effective weak disorder condition for the random masses. Notice that for weak disorder, the size of the frequency band does not change appreciably as compared to the non-disorder case; therefore, we can introduce a parameter $\mu$ that is related to the perturbed frequency $\omega$ through
\begin{eqnarray}
\omega^2=\frac{4K}{M}\sin^2\left(\frac{\mu a}{2}\right).
\label{disper} 
\end{eqnarray}
In the case of no disorder, $\mu$ is the wave number of a mode of the harmonic chain and Eq.~(\ref{disper}) is the dispersion relation. The second expression in Eq.~(\ref{weak}) establishes the condition of weak disorder for the random spring constants. From now on, $a$ will be set to one.

For our analysis, it is convenient to introduce the dimensionless random variables 
\begin{eqnarray*}
\widetilde{\delta m}_n=\frac{m_n-M}{M} \ \ \text{and} \ \  \widetilde{\delta k}_n=\frac{k_n-K}{K}, 
\end{eqnarray*}
which represent the normalized fluctuations of the random masses (or spring constants) around their mean values. In addition, if a quantity does not present the accent mark $\widetilde{\quad}$, it means that such variable is not normalized; i.e. $\delta m_n=m_n-M$.

As we will demonstrate below, our results will depend only on the following normalized binary correlators    
\begin{eqnarray}
\label{corre}
\chi_1(l)&=&\frac{\langle \delta m_n \delta m_{n-l} \rangle}{\sigma^2_m}, \nonumber \\
\chi_2(l)&=&\frac{\langle \delta k_n \delta k_{n-l} \rangle}{\sigma^2_k},  \\
\chi_3(l)&=&\frac{\langle \delta m_n \delta k_{n-l} \rangle}{\langle \delta k_n \delta m_n\rangle}. \nonumber
\end{eqnarray}
Here, we assume that disorder is spatially invariant in average; therefore, the binary correlators (\ref{corre}) depend only on the index difference $l$. We also assume
that $\chi_i(l)$ ($i=1,2,3$) is a decreasing and even function of $l$.

\section{The Hamiltonian map approach}
\label{Ham}

The Hamiltonian map approach is a method that maps the equations of the eigenstates for some one dimensional models into the phase space dynamics of a linear oscillator subjected to linear delta-kicks, where only the evolution  of the position $x_n$ and its conjugate momentum $p_n$ just after (or before) the $n$th kick are analyzed. Therefore, this dynamical approach implies that the eigenstate equations are solved as an initial value problem.

The Hamiltonian map approach has been widely used to obtain analytical estimations of $L_{\text{loc}}$ for the one dimensional Anderson model with diagonal disorder \cite{IKT95,IRT98}, the Kroning-Penney model with positional and structural disorder \cite{HITT10,IKU01}, harmonic chains with correlated mass disorder \cite{ZLW15}, electronic systems consisting of a set of potential barriers and/or wells with random heights \cite{HIM13}, among many other one dimensional systems (see Ref. \cite{IKM12} and the references therein for details on this subject).

In order to apply a perturbative approach, we take the Fourier transform of the dynamical equations (\ref{dynamic}) with respect to time, obtaining
\begin{equation}
m_n\omega^2 q_n=k_n \left[q_n-q_{n-1} \right]+k_{n+1}[q_n-q_{n+1}].
\label{fou}
\end{equation}
In the case of purely mass disorder, one can use the following transformation
\begin{eqnarray}
x_n=q_n, \quad  p_n=\frac{q_n\cos \mu-q_{n-1}}{\sin \mu},
\label{trans}
\end{eqnarray}
to map the equation of vibrational modes into the Hamiltonian map of a periodically kicked  linear oscillator, and therefore, the Hamiltonian map approach can be applied for this case. However, when one also considers spring disorder, it is no longer possible to establish a connection between the vibrational modes and the trajectories of the kicked oscillator, therefore, the Hamiltonian map approach can not be used. Nevertheless, if we still use transformation (\ref{trans}) in our present context, one can still develop a proper perturbation theory. Indeed, with the use of such transformation, instead of working directly with Eq.~(\ref{fou}), the problem is transformed into a two-dimensional map
\begin{eqnarray}
\frac{k_n}{k}\left(
                \begin{array}{r}
                x_{n+1}\\
                p_{n+1}
                \end{array}
         \right)=  \begin{array}{rr}
  
        \left(
                \begin{array}{rr}
                 a_{1,1} & a_{1,2}\\
                 a_{2,1} & a_{2,2} 
                \end{array}
         \right)\left(\begin{array}{r}
                x_n\\
                p_n
                \end{array}\right),
  \end{array}
\label{map}
\end{eqnarray}
with
\begin{eqnarray*}
a_{1,1}&=&\cos \mu-4\widetilde{\delta m}_n\sin^2\left(\frac{\mu}{2}\right)+2\widetilde{\delta k}_n \sin^ 2\left(\frac{\mu}{2}\right)+\widetilde{\delta k}_{n+1} \nonumber \\
a_{1,2}&=&\sin \mu+\widetilde{\delta k}_n\sin \mu \nonumber \\
a_{2,1}&=&-\sin \mu+\left[\left(\widetilde{\delta k}_n-2\widetilde{\delta m}_n\right) \cos \mu 
-\widetilde{\delta k}_{n+1}\right]\tan\left(\frac{\mu}{2}\right) \nonumber \\
a_{2,2}&=&\cos \mu + \widetilde{\delta k}_n \cos \mu.
\end{eqnarray*} 
In the case of no disorder, the matrix of the map (\ref{map}) represents a rotation with the angle $\mu$, and this is the key point in order to develop a perturbation theory as we will see below. Note that the condition of weak disorder (\ref{weak}) has not been used yet

It is convenient to introduce the  variables $(r_n,\theta_n)$ defined by the transformation $x_n=r_n\sin \theta_n$ and $p_n=r_n\cos \theta_n$. In this way, the map (\ref{map}) takes the form
\begin{eqnarray}
\frac{\cos \theta_{n+1}}{R_n}&=&\cos\left(\theta+\mu\right)+A\widetilde{\delta m}_n+B\widetilde{\delta k}_n+C\widetilde{\delta k}_{n+1} \nonumber \\
\frac{\sin \theta_{n+1}}{R_n}&=&\sin\left(\theta+\mu\right)+D\widetilde{\delta m}_n+E\widetilde{\delta k}_n+F\widetilde{\delta k}_{n+1}, \qquad
\label{tmap}
\end{eqnarray}
where
\begin{eqnarray*}
A&=&-2\tan\left(\frac{\mu}{2}\right)\cos\mu\sin \theta_n, \ \ R_n=\frac{K}{k_n}\frac{r_{n}}{r_{n+1}}\\
B&=&\cot\mu\left[\sin \theta_n-\sin(\theta_n-\mu) \right] \\
C&=&-\tan\left(\frac{\mu}{2}\right)\sin\theta_n, \ \ 
D=-4\sin^2 \left(\frac{\mu}{2}\right)\sin \theta_n \\
E&=&\sin \theta_n-\sin(\theta_n-\mu), \ \
F=\sin \theta_n.
\end{eqnarray*}
Now, making use of the weak-disorder conditions (\ref{weak}), the map
given by Eq. (\ref{tmap}) can be written, up to second order terms of the disorder, as
\begin{eqnarray}
\label{mapf}
\theta_{n+1}&=&\theta_n+\mu+2\widetilde{\delta m}_n\sin^ 2\theta_n\tan\left(\frac{\mu}{2}\right) 
 \\&+&\frac{\widetilde{\delta k}_{n+1} \cos\left(\theta_n+\mu/2\right)-\widetilde{\delta k}_n\cos\left(\theta_n-\mu/2\right)}{\cos\left(\mu/2\right)} \sin \theta_n\nonumber \\
&+&4\left(\widetilde{\delta m}_n\right)^2\tan^2\left(\frac{\mu}{2}\right)\sin^3\theta_n\cos\theta_n \nonumber \\
&-&\widetilde{\delta k}_n\widetilde{\delta k}_{n+1}\frac{\sin \theta_n\cos\left(\theta_n-\mu/2\right)\cos\left(2\theta_n+\mu/2\right)}{\cos^2\left(\mu/2\right)} \nonumber \\
&-&\frac{1}{2}\left(\widetilde{\delta k}_{n+1}\right)^2 \frac{\sin\left(2\theta_n+\mu\right)\sin^2\theta_n}{\cos^2\left(\mu/2\right)} \nonumber \\
&+&2\widetilde{\delta m}_n\widetilde{\delta k}_{n+1}\frac{\sin^2\theta_n\sin\left(\mu/2\right)\cos\left(2\theta_n+\mu/2\right)}{\cos^2\left(\mu/2\right)}\nonumber \\
&-&4\widetilde{\delta m}_n\widetilde{\delta k}_n\frac{\sin^2\theta_n\cos\theta_n\sin\left(\mu/2\right)\cos(\theta_n-\mu/2)}{\cos^2\left(\mu/2\right)} \nonumber \\
&+&\frac{1}{2}\left(\widetilde{\delta k}_{n}\right)^2 \frac{\cos^2\left(\theta_n-\mu/2\right)\sin(2\theta_n)}{\cos^2\left(\mu/2\right)} \ \ \ [\text{mod}\ 2\pi] . \nonumber
\end{eqnarray}

If $\mu$ is an irrational multiple of $\pi$, it sweeps quickly the whole interval $[0,2\pi]$ and, therefore, an uniform distribution $\rho(\theta)$ for the angle $\theta$ is observed:
\begin{equation}
\rho(\theta)\simeq \frac{1}{2\pi}.
\label{uni}
\end{equation}
When $\mu$ is a rational multiple of $\pi$, periodic orbits are formed in the map (\ref{mapf}) when no disorder is present. If disorder is switched on, the leading term of the distribution $\rho(\theta)$ can still be considered as uniform, but now some small modulations around $1/2\pi$ appear. However, our perturbative analysis breaks out  in the vicinity of $\mu=\pi$, as can be seen from the angular map (\ref{mapf}).   

A clarification needs to be done when $\mu\rightarrow0$: If one only considers mass disorder, then the dominant term of $\rho(\theta)$ is still given by Eq.~(\ref{uni}) because $\mu$ is still the leading term of the map (\ref{mapf}) even for this case. However, if spring disorder is also considered, we cannot apply the same argument to use Eq.~(\ref{uni}). Nevertheless, numerical simulations for the specific models analyzed in section \ref{Num}, together with some already published analytical results discussed in section \ref{locS},  show that Eq.~(\ref{uni}) is still  valid at least for some cases in the presence of spring disorder in the low frequency limit. 

There is another reason to believe that Eq. (\ref{uni}) is valid for any case: In Ref. \cite{A83}, a change of variables is done in Eq. (\ref{fou}) when purely bond disorder is considered. Under that transformation, the problem is reduced to study only diagonal disorder. In addition, if weak disorder is considered, the diagonal disorder corresponds to the springs constants $k_i$ which are gauged by $\omega^2$ just as in the case of purely mass disorder. Thus, if we use the Hamiltonian map approach after the aforementioned transformation, $\mu$ will be the dominant term, and therefore, Eq. (\ref{uni}) is valid. 

\section{The localization length}
\label{locS}

The inverse localization length (Lyapunov exponent $\lambda$) is defined by the expression
\begin{equation}
L^{-1}_{\text{loc}}\equiv\lambda=\lim_{N\rightarrow \infty}\frac{1}{N}\sum^N_{n=1} \ln \left|\frac{q_n}{q_{n-1}}\right| ,
\label{loc}
\end{equation} 
where the evolution of $q_n$ is defined in Eq. (\ref{fou}). In terms of the variables $r_n$ and $\theta_n$, Eq. (\ref{loc}) can be written as
\begin{equation}
\lambda= \lim_{N\rightarrow \infty}\frac{1}{N}\sum^N_{n=1} \ln \left(\frac{r_n}{r_{n-1}}\right)+\lim_{N\rightarrow \infty}\frac{1}{N}\sum^N_{n=1} \ln \left|\frac{\sin \theta_n}{\sin \theta_{n-1}}\right|.
\label{loc1}
\end{equation}
Nevertheless, for a flat distribution of angles the second term of the r.h.s of the latter equation vanishes, and therefore,
\begin{equation}
\lambda= \lim_{N\rightarrow \infty}\frac{1}{N}\sum^N_{n=1} \ln \left(\frac{r_n}{r_{n-1}}\right)=\left<\ln\left(\frac{r_n}{r_{n-1}}\right)\right>.
\label{loc2}
\end{equation}
Now, if one uses the two dimensional map (\ref{tmap}), Eq. (\ref{loc2}) can be rewritten in terms of the angle variable 
\begin{equation}
\lambda=-\frac{1}{2}\left< \ln\left|\frac{d\theta_{n+1}}{d\theta_n} \right| \right> .
\end{equation}
By introducing  the angular map (\ref{mapf}) into the latter expression and by keeping terms up to second order of the disorder, we obtain
\begin{eqnarray}
\lambda&=&-\tan\left(\frac{\mu}{2}\right)\langle\widetilde{\delta m}_n \sin(2\theta_n)\rangle+\frac{\langle\widetilde{\delta k}_n\cos\left(2\theta_n-\mu/2\right)\rangle}{2\cos\left(\frac{\mu}{2}\right)} \nonumber \\
&+&\frac{\widetilde{\sigma}^2_m}{2} \tan^2\left(\frac{\mu}{2}\right)-\frac{\langle\widetilde{\delta k}_{n+1}\cos\left(2\theta_n+\mu/2\right)\rangle}{2\cos\left(\frac{\mu}{2}\right)}+\frac{\widetilde{\sigma}^2_k}{4\cos^2\left(\frac{\mu}{2}\right)}\nonumber\\
&-&\tan^2\left(\frac{\mu}{2}\right)\frac{\widetilde{\Delta}}{2}\left[1+\chi_3(1)\right]-\widetilde{\sigma}^2_k\frac{\chi_2(1)\cos\mu}{4\cos^2\left(\frac{\mu}{2}\right)} . \nonumber\\
\label{loc3}
\end{eqnarray}
Here, the condition of uniform distribution (\ref{uni}) for the random variable $\theta_n$ has been used, and also, it has been introduced the following notation
\begin{equation}
\widetilde{\Delta}=\left<\widetilde{\delta m}_n \widetilde{\delta k}_n\right> . 
\end{equation}

In order to obtain the final expression for the localization length, we need to compute the binary correlators between the disorder variables and the trigonometric functions of the random variable $\theta_n$ that appear in Eq. (\ref{loc3}). This is done in Appendix \ref{appex}. Therefore, our expression for the inverse localization length reads as
\begin{equation}
\lambda=\frac{\tan^2\left(\frac{\mu}{2}\right)}{2}\left[\widetilde{\sigma}^2_mW_1(\mu)+\widetilde{\sigma}^2_kW_2(\mu)
+2\widetilde{\Delta}\cos\mu W_3(\mu)\right],
\label{final}
\end{equation}
where the functions $W_i(\mu)$ are defined through the expressions
\begin{eqnarray*}
W_1(\mu)=1+2\sum^{\infty}_{l=1}\chi_1(l)\cos(2l\mu) \\
W_2(\mu)=1+2\sum^{\infty}_{l=1}\chi_2(l)\cos(2l\mu) \\
W_3(\mu)=1+2\sum^{\infty}_{l=1}\chi_3(l)\cos(2l\mu).
\end{eqnarray*}
These expressions  correspond to the Fourier transforms of the normalized binary correlators $\chi_i(l)$ defined in Eq. (\ref{corre}). Therefore, formula (\ref{final}) is given by the sum of three terms; the first two terms describe the effects of purely mass and spring disorders, respectively, whereas, the third one represents the interplay between the two.  

Notice also that the functions $W_i$ are  even functions with period $\pi$ that satisfy the normalization condition
\begin{eqnarray}
\int^{\frac{\pi}{2}}_0 W_i(\mu)=\frac{\pi}{2} ,
\label{normal}
\end{eqnarray}
which follows from the fact that $\chi_i(0)=1$. In addition, the final expression for the inverse localization length (\ref{final}) can be expressed in terms of $\omega$ if one uses relation (\ref{disper}).

It is worthwhile to mention that formula (\ref{final}) is a generic expression since correlations of any kind are allowed for stationary sequences of random masses and spring constants. This formula also works for practically the whole frequency band with the exception of a small neighborhood at the right band edge ($\mu\simeq \pi$). In addition, the low frequency limit of $\lambda$ will depend on the particular model of harmonic chains with correlated disorder through the functions $W_i$. For example, in the case of uncorrelated disorder (i.e. $W_1=W_2=1$ and $W_3=0$) $\lambda\sim\omega^2$, or equivalently $L^{-1}_{\text{loc}}$ diverges as $\omega^{-2}$ in the low frequency limit; a result that is consistent with previous analytical estimations for harmonic chains with uncorrelated weak disorder (see Refs. \cite{LLP03,AOI18} and references therein).   

In the case of pure mass disorder, Eq. (\ref{final}) reduces to
\begin{eqnarray*}
\lambda=\frac{\widetilde{\sigma}^2_m}{2}\tan^2\left(\frac{\mu}{2}\right)\left[1+2\sum^{\infty}_{l=1}\chi_1(l)\cos(2l\mu)\right],
\end{eqnarray*}
which coincides with the expression derived in Ref. \cite{HIT10} obtained by direct comparison between the harmonic chain model and the one dimensional Anderson model. In addition, in the low frequency limit, the latter expression is equal to the one obtained by different means in Ref. \cite{ZLW15}.

By considering only mass and spring disorder with no correlations, Eq.~(\ref{final}) takes the form (in the low frequency limit) $\lambda\simeq[\widetilde{\sigma}^2_m+\widetilde{\sigma}^2_k]M\omega^2/8K$ which is equal to the expression given in Ref.~\cite{F21} that was obtained by using methods for dynamical systems. This supports the idea that the uniform distribution (\ref{uni}) is still valid in the low frequency limit even if spring disorder is present.

When dimeric correlations are considered in either mass or spring disorder, the normalized binary correlators take the form $\chi_{i}(l)=\delta_{l,0}+\delta_{|l|,1}/2$, $i=1,2$, $\widetilde{\Delta}=0$, and Eq.~(\ref{final}) reduces to
\begin{eqnarray*}
\lambda=\widetilde{\sigma}^2\tan^2\left(\frac{\mu}{2}\right)\cos{\mu} \ \ \text{with} \ \ \widetilde{\sigma}=\widetilde{\sigma}_m,\ \widetilde{\sigma}_k .
\end{eqnarray*}
This means that there is an extended vibrational mode at $\mu=\pi/2$ corresponding to the center of the frequency band. This is in complete agreement with the results presented in Refs.~\cite{DK94,DMS93}.

\section{Numerical results}
\label{Num}

In order to generate localization-delocalization transitions of vibrational modes with effective mobility edges $\mu_1$ and $\mu_2$, we propose the following  power spectra 
\begin{eqnarray}
W_1=W_2=
         \left\{
                \begin{array}{lll}
                  \frac{\pi}{2(\mu_2-\mu_1)} \  &\text{if} \  \mu \in [\mu_1,\mu_2] \cup [\pi-\mu_1,\pi-\mu_2]  \\ \\
                  0 \ \ &\text{otherwise} 
                \end{array}
         \right. ,
\label{case}
\end{eqnarray}
which satisfies the normalization condition (\ref{normal}) and $W_{1,2}(-\mu+\pi)=W_{1,2}(\mu)$ (this is a property that follows from the fact that $W_{1,2}$ are even functions with period $\pi$). Therefore, $\pi-\mu_1$ and $\pi-\mu_2$ are also effective  mobility edges. The power spectra (\ref{case}) correspond to self-long-range correlations of the form
\begin{eqnarray*}
\chi_1(l)=\chi_2(l)=\frac{1}{2(\mu_2-\mu_1)l}\left[\sin(2\mu_2 l)-\sin(2\mu_1 l) \right].
\end{eqnarray*} 

Now we define $W_3(\mu)$ as follows.
We use the methods presented in Ref.~\cite{HITT10} to generate two random sequences $\{m_i\}$ and $\{k_i\}$ that satisfy relation (\ref{case}) and the following expression for $W_3(\mu)$ 
\begin{equation}
W_3(\mu)=\widetilde{\sigma}_m\widetilde{\sigma}_k\sqrt{W_1(\mu)W_2(\mu)}\sin(2\delta),
\end{equation}
where the parameter $\delta$ lies in the interval $[-\pi/4,\pi/4]$ and controls the degree of correlation between the random variables $m_i$ and $k_i$: For $\delta=0$  no cross-correlations exist. Therefore, for this particular construction  of the random sequences,  Eq.~(\ref{final}) reads
\begin{eqnarray}
\lambda&=&\frac{\tan^2\left(\frac{\mu}{2}\right)}{2}\Big[\widetilde{\sigma}^2_mW_1(\mu)+\widetilde{\sigma}^2_kW_2(\mu) \nonumber  \\ 
&+& 2\widetilde{\sigma}_m\widetilde{\sigma}_k\sqrt{W_1(\mu)W_2(\mu)}\sin(2\delta)\cos\mu \Big].
\label{final1}
\end{eqnarray}

\begin{figure}[!t]
\includegraphics[width=8.6cm,height=7.4cm]{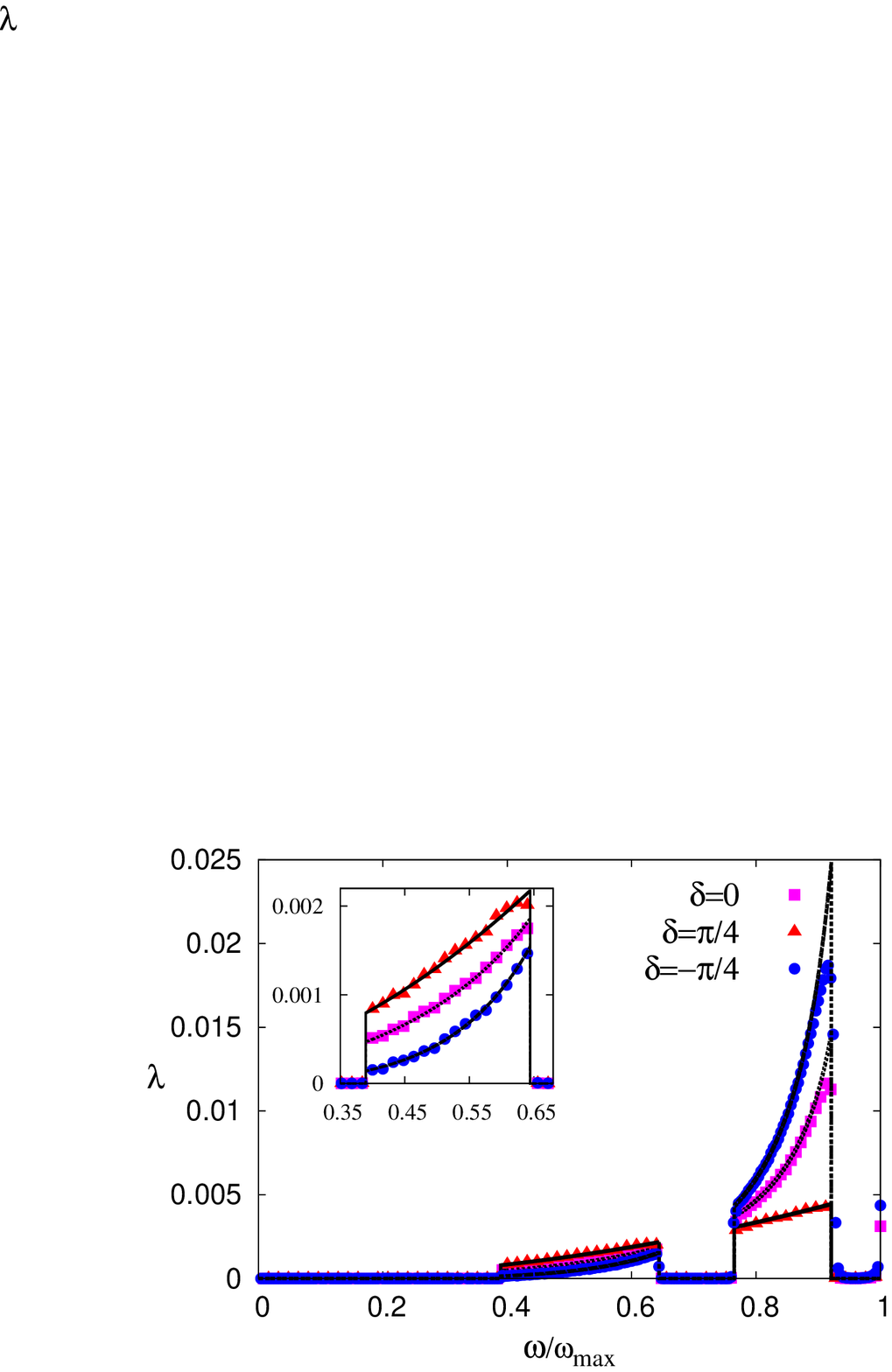}
\caption{Inverse localization length as a function of the normalized frequency for Case I. Symbols represent numerical simulations for a chain of length $N=2\times10^6$, whereas continuous, doted and dashed black lines correspond to the theoretical result (\ref{final1}) for $\delta=\pi/4$, $\delta=0$ and $\delta=-\pi/4$, respectively. The inset is an enlargement around the first localization window}
\label{Fig1}
\end{figure}

In our numerical simulations we analyze two cases: Case I corresponds to $\mu_1=0.8$ and $\mu_2=1.4$, whereas in Case II we have $\mu_1=0$ and $\mu_2=0.8$. In both cases  $\widetilde{\sigma}^2_m=\widetilde{\sigma}^2_k=0.001$ and $\omega_{\text{max}}=2\sqrt{K/M}$ is the maximum frequency of the allowed band. In this way, Case I produces two localization windows and three transparent ones as it is shown in Fig.~\ref{Fig1}, where there is a good correspondence between numerics and the analytical prediction~(\ref{final1}). In this figure, it is also compared the case without cross correlations ($\delta=0$) with the two extreme cases of positive ($\delta = \pi/4$) and negative ($\delta = -\pi/4$) cross correlations: In the first localization window, $\delta=-\pi/4$ decreases the inverse localization length, whereas, for $\delta=\pi/4$, $\lambda$ is increased. We have the opposite situation in the second localization window.

\begin{figure}[!t]
\includegraphics[width=8.6cm,height=7.4cm]{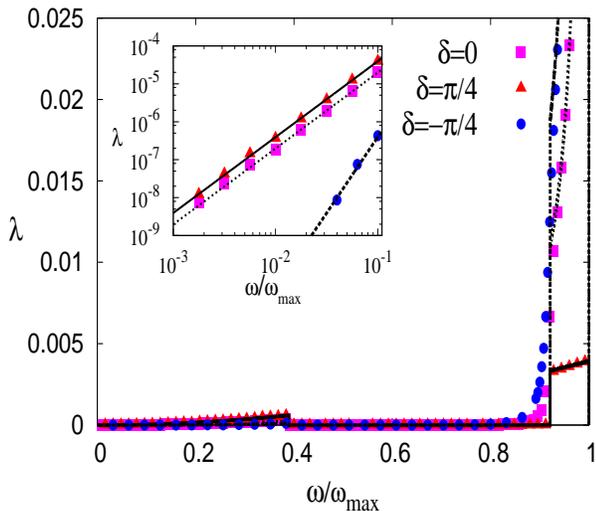}
\caption{Inverse localization length as a function of the normalized frequency for case II. Symbols represent numerical simulations for a chain of length $N=2*10^6$, whereas continuous, doted and black lines correspond to the theoretical result (\ref{final1}) for $\delta=\pi/4$, $\delta=0$ and $\delta=-\pi/4$, respectively. The inset shows $\lambda$ in log-scale in the low frequency regime for $N=2\times10^9$.}
\label{Fig2}
\end{figure}

\begin{figure}[!t]
\includegraphics[scale=0.63]{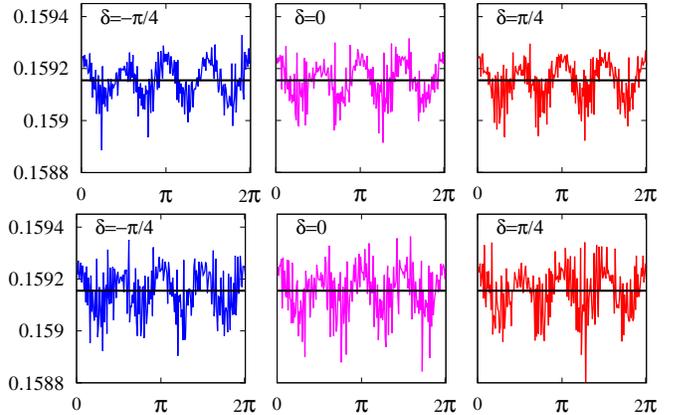}
\caption{Probability phase distribution functions $\rho(\theta)$ obtained after $5\times 10^8$ iterations of the map (\ref{tmap}) with $\mu=5\times10^{-5}$. The first row corresponds to Case I, whereas the second row represents Case II. The horizontal black lines are the uniform distribution $\rho(\theta)=1/2\pi$.}
\label{Fig3a}
\end{figure}

On the other hand Case II also produces two localization windows but only one transparent window as Fig. \ref{Fig2} shows. We also observe good correspondence between the analytical prediction (\ref{final1}) and numerical simulations. This good correspondence also holds in the low frequency regime as can be seen from the inset of the figure where $\lambda \sim \omega^2$ (or $L_{\text{loc}} \sim \omega^{-2}$, see Eq. (\ref{loc}))  for $\delta=0$ and $\delta=\pi/4$, whereas $\lambda \sim \omega^4$ ($L_{\text{loc}} \sim \omega^{-4}$)  for $\delta=-\pi/4$. However, values of $\mu$ much smaller than $\widetilde{\sigma}^2_k$ are needed in order to support the assumption that Eq. (\ref{uni}) is valid even in the low frequency regime when bond disorder is present. Since smaller values of $\mu$ for the calculation of $\lambda$ exceed the practical time of simulation, we decided to include another figure: In Fig. \ref{Fig3a}, we plot six different probability distribution functions $\rho(\theta)$ corresponding to different values of $\delta$ for Cases I and II. As one can clearly see, $\rho(\theta)$ is practically flat with small modulations around the value $1/2\pi$ and the parameter $\mu$ is small as compared with $\widetilde{\sigma}^2_k$ ($\mu$ is twenty times smaller than $\widetilde{\sigma}^2_k$). This  supports the assumption of the uniform phase distribution (\ref{uni}) when spring disorder is present in the low frequency limit, where the argument of the rapid dynamics of the angle $\mu$ can not be used.

\begin{figure}[!t]
\includegraphics[width=8.6cm,height=7.4cm]{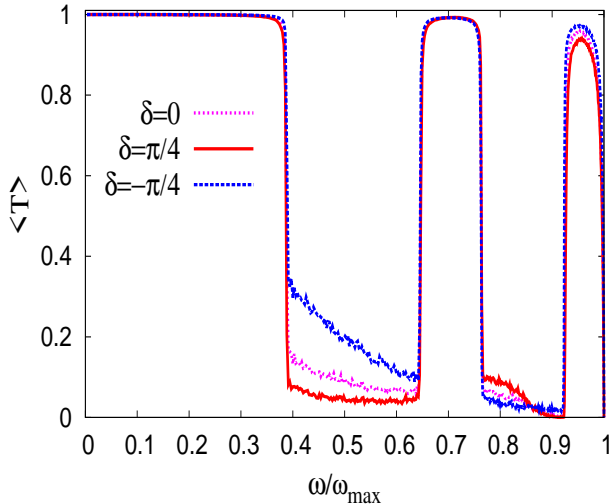}
\caption{Transmission coefficient averaged over 1000 chain realizations as a function of the normalized frequency for Case I. Numerical simulations are done for chains of length $N=1000$.}
\label{Fig3}
\end{figure}

\begin{figure}[!t]
\includegraphics[width=8.6cm,height=7.4cm]{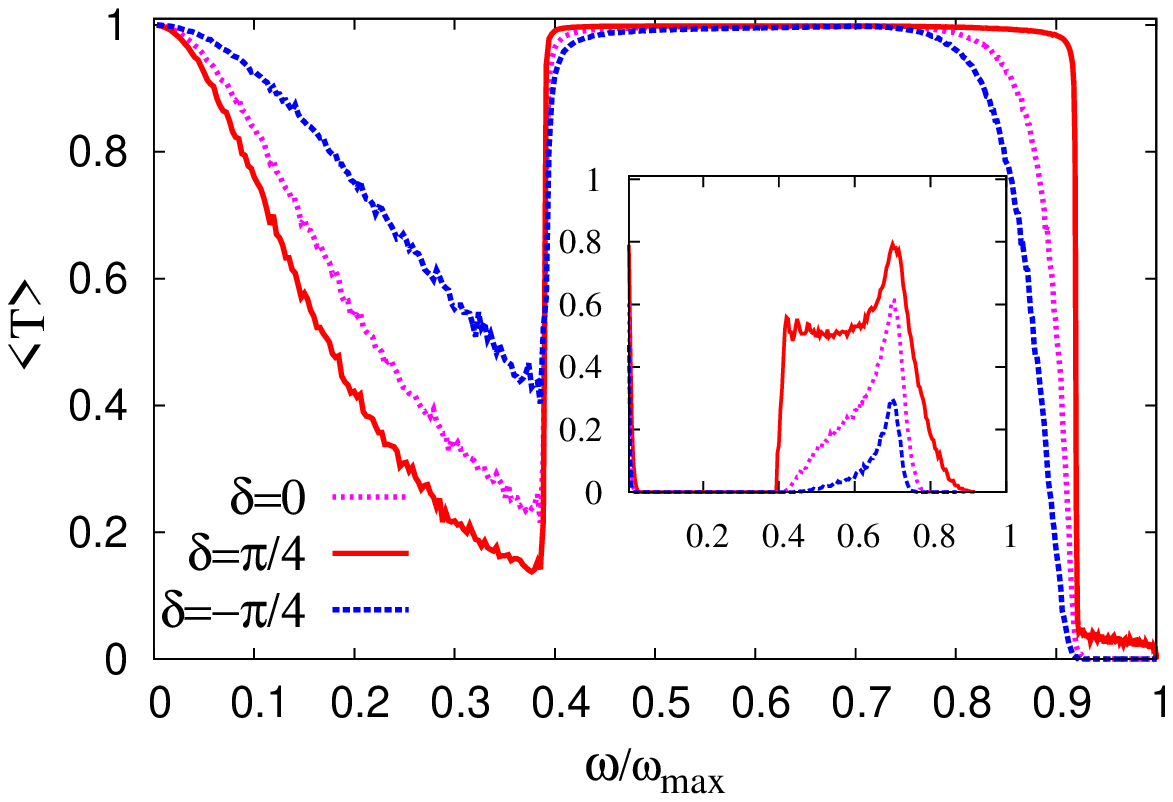}
\caption{Transmission coefficient averaged over 1000 chain realizations as a function of the normalized frequency for Case II. Numerical simulations are done for chains of length $N=1000$ (main panel) and $N=2\times10^6$ (inset). }
\label{Fig4}
\end{figure}

Now, we study the transport properties of finite disordered harmonic chains embedded between two reservoirs consisting on two semi-infinity homogeneous chains of masses $M$ and spring constants $K$. The transport properties are determined by the length of the chain $N$ and the localization length $L_{\text{loc}}$. In particular, we are interested in the transmission coefficient that we compute using the transfer matrix approach (see the supplemental material in Ref. \cite{AABOI19}). We clearly see that the localization windows and transparent windows manifest themselves on the average transmission coefficient $\langle T\rangle$ as Fig. \ref{Fig3} and Fig. \ref{Fig4} show for Case I and Case II, respectively. For the chain length shown in the figures, the localization windows inhibits the transport of phonons, as expected, and in the transparent windows $\langle T\rangle$ remains close to one. The reason why $\langle T\rangle \ne 1$ in the transparent windows relies on the fact that $\lambda$ is zero only within the second order of the perturbative approach. Therefore, as the system size becomes larger the transmission coefficient of the modes belonging to the transparent windows will decrease (see the inset of Fig. \ref{Fig4}), and eventually  for a sufficiently large chain length these effective transparent windows will disappear. It is relevant to note that while the mobility edges are determined exclusively from the self-correlations, cross-correlations diminish or enhance the transmission.     

\section{Thermal conductivity}
\label{conducs}

It is important to relate the results of the previous section to the problem of heat conduction through disordered harmonic chains. For this purpose, we consider that the system  is embedded between two thermal reservoirs formed by an infinite set of equal masses $m$ and spring constants $k$, which define the so-called oscillators baths. In addition, the expression for the stationary heat flux through the harmonic chain can be written in the classical limit as \cite{D08}
\begin{eqnarray}
J=\frac{k_B\Delta T}{2\pi}\int^{\infty}_0 T(\omega)d\omega.
\label{conduc}
\end{eqnarray}       
Here, $T(\omega)$ is the transmission coefficient discussed in the previous section, $k_B$ is the Boltzmann constant, and $\Delta T$ is the temperature difference between the heat reservoirs. Therefore, when there are mobility edges, the contribution to the heat flux $J$ from modes belonging to localization windows can be neglected for sufficient large chain sizes. In this way, one may construct a thermal filter where the contribution to heat flux only comes from modes belonging to  transparent windows. In addition, from Eqs. (\ref{Fou}) and (\ref{conduc}) we can see that the thermal conductivity  scales as
\begin{equation}
\kappa \sim N,
\label{thermal}
\end{equation}
which is the same scaling behavior observed in homogeneous chains (no disorder) \cite{RLL67}.

In the present context, we have mobility edges within the second order approximation, this means that the ballistic behavior predicted by equation (\ref{thermal}) will be valid over a wide range of chains lengths. Nevertheless, for large enough chains, the effective transparent windows will be destroyed and the true asymptotic behavior of $\kappa$ will depend only on the modes whose frequencies are close to zero.

If an effective transparent window contains the zero frequency mode (Case I), then the next higher order term of the inverse localization length is needed to determine the size scaling of $\kappa$. On the other hand, when an effective transparent window  does not contain the zero frequency mode (Case II), and for large enough chain sizes, the scaling behavior of $k$ is determined by a finite frequency window of vibrational modes near the zero frequency mode. In this window modes are extended and practically equal to those of the homogeneous chain, therefore, the transmission coefficient must also be practically equal to the transmission coefficient of the ordered chain $T^{\mbox o}$; this argument has been used extensively in the literature which is also supported by numerical results (see for example Refs. \cite{HM20,D01,RD08}). In this way, $J$ can be written as
\begin{equation}
J\sim \int^{\omega_c}_0 T^{\mbox o}(\omega) d\omega,
\label{int}
\end{equation}   
where $\omega_c \ll 1$ is a cut-off frequency that depends on the localization length and the chain size through the equation
\begin{equation}
L_{\text{loc}}(\omega_c)=N.
\label{cut}
\end{equation}
In addition, since there exist analytical expressions for $T^{\mbox o}$ (see Ref. \cite{RD08J}), one can compute the integral (\ref{int}) in the low frequency limit obtaining (see Eq.~(16) of Ref. \cite{HM20})
\begin{eqnarray}
  J\sim
         \left\{
                \begin{array}{ll}
                {\displaystyle
                   \omega^3_c(N) } & \ \mbox{for fixed BC},  \vspace{0.2cm} \\     
                  {\displaystyle            
                   \omega_c(N) } & \ \mbox{for free BC} . 
                \end{array}
         \right. 
\label{a2} 
\end{eqnarray} 
Here BC is an abbreviation for boundary conditions. Therefore, the low frequency behavior of the inverse localization length (\ref{final}) together with Eqs. (\ref{Fou}), (\ref{cut}) and (\ref{a2}) determine how the thermal conductivity scales with the system size as long as all the non-zero frequency  modes become localized in the thermodynamic limit.

From the above discussion we conclude that for Case II with $\delta=0$ and $\delta=\pi/4$, where $L_{\text{loc}}\sim \omega^{-2}$,  $\omega_c\sim 1/\sqrt{N}$, and
\begin{eqnarray}
\kappa\sim
         \left\{
                \begin{array}{ll}
                {\displaystyle
                   N^{-1/2} } & \ \mbox{for fixed BC},  \vspace{0.2cm} \\     
                  {\displaystyle            
                    N^{1/2}} & \ \mbox{for free BC} ;
                \end{array}
         \right.
\label{a4}          
\end{eqnarray}  
as it should be since there are other disorder models where $L_{\text{loc}}\sim \omega^{-2}$ and Eq. (\ref{a4}) is also obtained \cite{AOI18,I73}. 

For Case II with $\delta=-\pi/4$, where $L_{\text{loc}}\sim \omega^{-4}$, we can use similar arguments yielding to the following result for the thermal conductivity 
\begin{eqnarray}
\kappa\sim
         \left\{
                \begin{array}{ll}
                {\displaystyle
                   N^{1/4} } & \ \mbox{for fixed BC},  \vspace{0.2cm} \\     
                  {\displaystyle            
                    N^{3/4}} & \ \mbox{for free BC} . 
                \end{array}
         \right.
\label{a5}          
\end{eqnarray}
Therefore, for both boundary conditions the thermal conductivity diverges in this case. 

\begin{figure}[!t]
\centering
\includegraphics[width=0.483\textwidth]{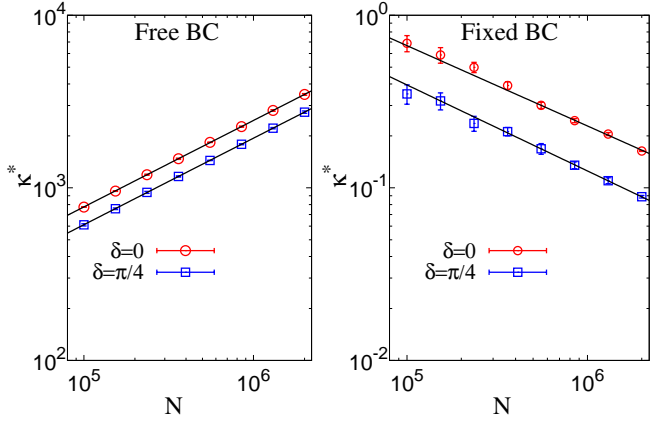}
\caption{$\kappa^*$ averaged over 10 disorder realizations as a function of the chain length $N$ for free and fixed boundary conditions (symbols). Black continuous lines are the best fits of the data with the function $f(N)=aN^b$. For free BC we got $b=0.501 \pm 0.001$ ($\delta=0$) and $b=0.502 \pm 0.004$ ($\delta=\pi/4$), whereas for fixed BC we obtained $b=-0.48\pm 0.03$ ($\delta=0$) and $b=-0.498 \pm 0.005$ ($\delta=\pi/4$). For free BC $\widetilde{\sigma}^2_m=\widetilde{\sigma}^2_k=0.001$, while for fixed BC $\widetilde{\sigma}^2_m=\widetilde{\sigma}^2_k=0.003$. $k_B$ was set to one.}
\label{Fig5}
\end{figure}

It is pertinent to notice from the inset of Fig. \ref{Fig4} that the transmission coefficient is still not negligible for modes belonging to the effective transparent window even for a chain length of $N=2\times10^6$.  This is a problem when trying to validate predictions (\ref{a4}) and (\ref{a5}) with numerical simulations because these predictions only take into account the contribution of low frequency modes to heat conduction. Although, the effective transparent windows will disappear for larger chain lengths, in our numerical simulations we can only reach values around $N=2\times10^6$. To overcome this issue,  we define $J^{*}$ as the heat flux from all the modes not belonging to the transparent window, consequently, we define $k^{*}$ as
\begin{equation}
\kappa^{*}=\frac{J^{*}N}{\Delta T}.
\label{Fou2}
\end{equation}
In this way we should be able to observe the thermal conductivity scaling due to the contribution of the low frequency modes only. Indeed, in Fig. \ref{Fig5} we observe an excellent agreement between prediction (\ref{a4}) and numerical simulations.

Unfortunately, we were not able to validate prediction (\ref{a5}) with numerical simulations due to the fact that the inverse localization length is too small in the low frequency regime for Case II with $\widetilde{\sigma}_m=\widetilde{\sigma}_k$ and $\delta=-\pi/4$, and the asymptotic scaling behavior of $\kappa$ can only be observed for $N \gg L_{\text{loc}}$. Indeed, as can be seen in the inset of Fig. \ref{Fig2}, the low frequency asymptotic behavior of $L^{-1}_{\text{loc}}$ starts around $6\times10^{-7}$; therefore we need $N\gg1.7\times10^6$, which exceeds our numerical capabilities. We also tried to turn around this issue by incrementing the disorder intensity, but the second condition of Eq. (\ref{weak}) represents a severe restriction when trying to decrease the localization length using this approach with $\widetilde{\sigma}_m=\widetilde{\sigma}_k$. In fact we made simulations of heat conduction with $\widetilde{\sigma}^2_m=\widetilde{\sigma}^2_k=0.004$ and the asymptotic behavior (\ref{a5}) was still not observed; while for values beyond $0.004$ we observe that our perturbative approach starts to fail.

Finally, it is important to stress that the thermal conductivity cannot be measured directly in real experiments; instead one measures the thermal conductance $G$, which  is related to $\kappa$ through $G=\kappa/N$. 

\section{Conclusion}
\label{con}

In the present study, we have analyzed the localization and transport properties of harmonic chains with weak mass and spring disorder, where all correlations are allowed: mass disorder correlations, spring disorder correlations, and mass-spring disorder correlations. We obtained an expression for the inverse localization length within a second order of a perturbative approach, see Eq. (\ref{final}); which is valid for practically the whole frequency band, except in a vicinity of the right band edge. In the low frequency limit, there is no rigorous proof for Eq. (\ref{final}) if spring disorder is present; however, numerical simulations and some previous analytical results show that our expression works reasonably well. 


Equation (\ref{final}) allow us to produce and to manipulate transparent frequency windows which modulate directly the transport properties of the chain: Effective mobility edges are created with the self-correlations of the disorder, while cross-correlations are used to fine-tune the transmission coefficient.

Finally, we obtained the size scaling of the thermal conductivity from the expression of the localization length, see Eqs (\ref{a4}) and (\ref{a5}). The relevance of the role of the mobility edges to the heat conduction was also addressed.

We believe that our results and predictions could be experimentally verified and/or implemented in technological applications.

\appendix 

\section{}
\label{appex}

The binary correlators appearing in Eq. (\ref{loc3}) can be computed by introducing the following quantities 
\begin{eqnarray*}
J_l&=&\left<\widetilde{\delta m}_n \text{e}^{2i\theta_{n-l}} \right>,\ \ H_l=\left<\widetilde{\delta k}_n \text{e}^{2i\theta_{n-l}} \right>, \\
Z_l&=&\left<\widetilde{\delta k}_{n+1} \text{e}^{2i\theta_{n-l}} \right>, 
\end{eqnarray*}
here, $i$ is the imaginary unit.

In what follows, we explain in detail how to obtain $\left<\widetilde{\delta m}_n \cos(2\theta_n)\right>$ and $\left<\widetilde{\delta m}_n \sin(2\theta_n)\right>$ from $J_l$. The other binary correlators can be obtained in a similar manner from $H_l$ and $Z_l$, therefore only their final expressions will be given.

From the expression for $J_l$, we write
$$
J_{l-1}=\left<\widetilde{\delta m}_n \text{e}^{2i\theta_{n+1-l}} \right>.
$$
If we substitute Eq. (\ref{mapf}) into the latter expression, expand in Taylor series the complex exponential function, and  keep terms up to second order of the disorder, we get
\begin{eqnarray*}
\text{e}^{-2i\mu}q_{l-1}&=&q_l+4i\tan\left(\frac{\mu}{2}\right)\widetilde{\sigma}^2_m\chi_1(l)\left<F(\theta_{n-l})\right> \\
&-&2i\widetilde{\Delta}\frac{\chi_3(l+1)}{\sin\mu}\left[\left<F(\theta_{n-l})\right>-\left<L(\theta_{n-l})\right>\right] \\
&+&2i\widetilde{\Delta}\frac{\chi_3(l)}{\cos\left(\frac{\mu}{2}\right)}\left<G(\theta_{n-l})\right>,
\end{eqnarray*}
where 
\begin{eqnarray*}
F(\theta_{n-l})&=&\text{e}^{2i\theta_{n-l}}\sin^2\theta_{n-l},\\ 
G(\theta_{n-l})&=&\text{e}^{2i\theta_{n-l}}\sin\theta_{n-l}\cos\left(\theta_{n-l}+\mu/2 \right),\\
L(\theta_{n-l})&=&\text{e}^{2i\theta_{n-l}}\sin(\theta_{n-l}-\mu)\sin\theta_{n-l}.
\end{eqnarray*}
The disorder average of the latter functions that involves only the angle $\theta_{n-l}$ can be computed using the condition (\ref{uni}), then after some algebra we get
\begin{eqnarray*}
J_{l-1}\text{e}^{2i\mu(l-1)}&=&\text{e}^{2i\mu l}J_l-i\tan\left(\frac{\mu}{2}\right)\text{e}^{2i\mu l}\widetilde{\sigma}_m\chi_1(l)  \\
&+&\frac{i\text{e}^{2i\mu l}}{2\sin \mu}\widetilde{\Delta}\chi_3(l+1)\left[1-\text{e}^{i\mu} \right] \\
&-&\frac{\text{e}^{2i\mu l}}{2\cos\left(\frac{\mu}{2}\right)}\widetilde{\Delta}\chi_3(l)\text{e}^{-i\mu/2} .
\end{eqnarray*}
If we now take the sum from $0$ to $\infty$ on both sides of the last equation, after some algebraic manipulations, the following equation is obtained
\begin{eqnarray}
\text{Im}\left(J_0\right)&=&-\Bigg[\widetilde{\Delta}\left(\frac{\chi_3(1)}{2}+\sum^{\infty}_{l=1}\chi_3(l)\cos[\mu(2l-1)] \right) \nonumber \\
&+&\widetilde{\sigma}^2_m\sum^{\infty}_{l=1}\chi_1(l)\cos(2\mu l)\Bigg]\tan\left(\frac{\mu}{2}\right) .
\label{A1}
\end{eqnarray}

With a similar procedure, the following formulas are obtained:
\begin{eqnarray}
H_0&=&\widetilde{\sigma}^2_k\Bigg[\frac{1}{2}-\cos\mu-i\tan\left(\frac{\mu}{2}\right)\Bigg(\sum^{\infty}_{l=1}\chi_2(l)\text{e}^{i\mu(2l+1)}+\frac{1}{2}\nonumber \\
&+&\cos\mu\Bigg)\Bigg]-i\widetilde{\Delta}\tan\left(\frac{\mu}{2}\right)\text{e}^{i2\mu}\left[1+\sum^{\infty}_{l=1}\chi_3(l)\text{e}^{i2\mu l} \right]\nonumber \\
\label{A2}
\end{eqnarray}
and
\begin{eqnarray}
Z_0&=&\widetilde{\sigma}^2_{k}\left[i\tan\left(\frac{\mu}{2}\right)\left(\sum^{\infty}_{l=1}\chi_2(l)\text{e}^{i\mu(2l-1)}+\frac{1}{2}\right)+\frac{1}{2}\right]\nonumber \\
&-&i\tan\left(\frac{\mu}{2}\right)\widetilde{\Delta}\sum^{\infty}_{l=0}\chi_3(l)\text{e}^{2i\mu l} .
\label{A3}
\end{eqnarray}

Notice  the following identities: $\text{Im}\left(J_0\right)=\left<\delta m_n\sin2\theta_n\right>$, 
$\text{Re}\left(H_0\right)=\left<\delta k_n\cos2\theta_n\right>$,
$\text{Im}\left(H_0\right)=\left<\delta k_n\sin2\theta_n\right>$, and so on. 
Therefore, by replacing Eqs. (\ref{A1})-(\ref{A3}) into Eq. (\ref{loc3}), Eq. (\ref{final}) is obtained after some algebra.

\begin{acknowledgments}
J. A. M. -B. thanks support from CONACyT (Grant No. 286633), CONACyT-Fronteras (Grant No. 425854),
and VIEP-BUAP (Grant No. 100405811-VIEP2022), Mexico.
I. F. Herrera-Gonz\'alez thanks to professor Luca Tessieri for useful discussion and comments.
\end{acknowledgments}



\end{document}